\documentclass{aa}
\usepackage{amsmath}
\usepackage{natbib}
\bibpunct{(}{)}{;}{a}{}{,}
 \usepackage{txfonts}
\usepackage{graphicx}
\usepackage{color}
\title{Nanoflare statistics in an active region 3D MHD coronal model}
\author{S.~Bingert
        \and
        H.~Peter
        }
\institute{Max Planck Institute for Solar System Research (MPS),
           37191 Katlenburg-Lindau, Germany, [bingert,peter]@mps.mpg.de
           }

\usepackage{hyperref}
\hypersetup{
  pdfauthor = {Sven Bingert},
  pdfkeywords = {MHD,Pencil-Code,Corona}
  pdfsubject = {Sammlung von Ideen},
  pdfkeywords = {MHD,Pencil-Code,Corona}
}
\date{Received ... / Accepted ...}
  \abstract
  {}
  {We investigate the statistics for the spatial and temporal distribution
        of the energy input into the corona in a
    three-dimensional magneto-hydrodynamical (3D MHD) model. The model
    describes the temporal evolution of the corona above an observed active
    region. The model is driven by photospheric granular motions that braid
        the magnetic field lines. This induces currents that are dissipated,
        thereby leading to transient heating of the coronal
    plasma. We evaluate the transient heating as subsequent heating events and
        analyze their statistics. The results are then interpreted in the context of
    observed flare statistics and coronal heating mechanisms.
    Observed solar flares and other smaller transients cover a wide range of energies. The frequency
    distribution of energies follow a power law, the lower end of the
    distribution given by the detection limit of current
    instrumentation. One particular heating mechanism is based on the occurrence
    of so-called nanoflares, i.e. very low-energy deposition events.
  }
  {To conduct the numerical experiment we use a high-order finite-difference
    code that solves the partial differential equations for the
    conservation of mass, the momentum and energy balance, and the induction
    equation.  The energy balance includes Spitzer heat conduction and
    optically thin radiative losses in the corona. }
  {The temporal and spatial distribution of the Ohmic heating in the 3D MHD model
    follows a power law and can therefore be understood as a system in a
    self-organized critical state. The slopes of the power law are similar to the results
        based on observations of flares and smaller transients. We find that the
    coronal heating is dominated by
     events similar to the so-called nanoflares with energies on the order of 10$^{17}$\,J or 10$^{24}$\,erg.
  }
  {}
\keywords{Sun:corona
          --- Stars: coronae
          --- Magnetohydrodynamics (MHD)
          --- Methods: numerical}
\begin{document}

 \maketitle
\section{Introduction}

In the upper atmosphere of the Sun and stars, energy is released in a
transient fashion. Strong flares are just the tip of the iceberg, where the
plasma can be heated up to 10 MK or more and particles are accelerated to
relativistic speeds. More numerous are smaller energy releases that cause
smaller brightenings in X-rays and extreme UV, down to the current detection
limit. There is a continuous connection from these smallest observable
brightenings, often called microflares, to the largest of flares in the
sense that the distribution in energy of all these events roughly follows a
power law. Theoretical arguments have been proposed that even smaller
transient releases of energy have to exist, the nanoflares
\citep{1988ApJ...330..474P}, which could play a major role in the heating of
the corona. In this study we investigate the energy distribution of these
low energy releases as they are found in 3D magnetohydrodynamics (MHD)
models.

\cite{Rosner1978a} have already investigated the transient events of the
Sun, flaring stars, and other X-ray sources. They found that the energy
distribution of these events roughly follows a power law. To explain the
flare statistics \cite{Rosner1978a} provided a theoretical model that led to
a power-law distribution of the flare parameters. This model was based on
the assumptions that flaring is a stochastic relaxation, the waiting time
between two events is an independent parameter, and the energy release
during a flare is large in comparison to the energy content after the
flare. The power-law nature is not only observable for X-ray sources, but
also solar radio bursts show such a distribution as already studied by
\cite{Akabane:1956}.

These results were followed by numerous observations of solar X-ray flares
\citep{Crosby+al:1993,Crosby+al:1998,Christe+al:2008a}.  They have found
that not only the flare energy but also other flare parameters, such as the
flare peak flux or flare duration, follow a power law. A review of the
statistics of flares and their small counterparts, the microflares, is given
by \cite{Hannah+al:2011}.

To understand the contribution to the heating of the hot coronal plasma,
interpreting the power law is crucial. Following the observation-based
results, the power law has a negative exponent: small flares are more
numerous than large flares.  If the frequency of small flares is high enough
(the power law slope being steeper than $-$2), the heating would be
dominated by small events, the nanoflares. These have been proposed by
\cite{1988ApJ...330..474P} based on theoretical arguments and are not
directly observable with current instrumentation. \cite{1988ApJ...330..474P}
estimated an energy of roughly 10$^{17}$\,J (10$^{24}$\,erg) and named those
nanoflares because they are smaller than the convention of a microflare. The
latter have energies on the order of 10$^{20}$\,J and thus about 10$^{-6}$
times the energy of a large flare.  To understand how much nanoflares
contribute to the heating or if larger flares are composed of smaller
flares, it is important to know the distribution of the energies of these
transient events down to the lowest energy levels.

The observational approach is hereby limited by a lower threshold given by
the detection limit of the observation.  In addition, the flare energy is
only a derived quantity inferred from the flux of exteme UV and/or X-ray
photons, and is therefore only an indicator for the actual energy
input. Energy is also lost through other processes, such as heat conduction
or particle acceleration, and the relative merit of this energy loss
mechanism might vary with the total energy of the respective
event. Therefore even the power law index for the distribution of the actual
energy input and for the observed energy loss might differ (for one given
channel, e.g., the X-ray flux).

An alternative approach is the numerical modeling of these flare
statistics. They are based on the theoretical assumptions leading to a power
law distribution. A famous model uses the assumption that the system, e.g.,
the solar corona, is in a self-organized critical state
\citep{Lu&Hamilton:1991}. The concept of the self-organized criticality was
introduced by \cite{Bak+al:1987}. Small changes to the system will trigger
further reactions that can be compared to avalanches
\citep{Charbonneau+al:2001}. But the question whether the coronal magnetic
field is in such a state remains open \citep{Lu:1995a}.

Another numerical approach has been employed recently by
\cite{Dimitropoulou+al:2011}. They used the cellular automaton model to
produce an average power law index of -1.80 for the peak energies. For that
study they used several different observed active regions. However, in their
model the magnetic field boundary remains constant over the simulation time,
and the driving is done artificially, but both are not realistic.

In this study we present for the first time the frequency distribution of
transient heating events in a forward model approach that is based on a 3D
MHD model of a solar active region. The hot corona in this model is
sustained by Ohmic dissipation of currents, which result from the braiding
of magnetic field lines by convective motions prescribed at the lower
boundary.  Since the first of this type of forward model
\citep{Gudiksen+Nordlund:2002,Gudiksen+Nordlund:2005a,Gudiksen+Nordlund:2005b},
it has been shown that these models not only produce a hot loop-dominated
corona, but that they also match the (averaged) properties of spectral EUV
observations \citep{Peter+al:2004,Peter+al:2006}. Furthermore, the synthetic
observations show a comparable temporal variability \citep{Peter:2007}, and
new suggestions have been made on the nature of the observed net Doppler
shifts \citep{Peter+Judge:1999} in the transition region and corona
\citep{Hansteen+al:2010,Zacharias+al:2011.doppler}. These 3D MHD coronal
models show that the heating is concentrated in field-aligned current
structures and that they feature a strong temporal variability
\citep{2011A&A...530A.112B}. The spatio-temporal distribution of the heat
input also gives rise to individual coronal loops that appear in
(synthesized) extreme UV emission as having a constant cross section
\citep{Peter+Bingert:2012}, just as in observations.

The success of previous models provides evidence that the distribution of
the heating rate in space and time (on the resolved scales) is close to what
we find on the Sun. Therefore it is timely to investigate details of the
statistics of the energy deposition in the 3D MHD models, in particular the
energy distribution of the heating events, which is the goal of this study.

In Sect.~\ref{sec:model} we briefly describe the 3D MHD coronal model and
show in Sect.~\ref{sec:temp_point} the transient nature of the
self-consistent coronal heat input. In Sect.~\ref{sec:stat} we derive the
energy distribution of the heating events and discuss in Sect.~\ref{sec:obs}
some observational consequences, before we investigate some aspects of
self-organized criticality in our system in Sect~\ref{sec:disc_self_sim}.

\section{Coronal model} \label{sec:model}

\begin{figure}
    \resizebox{\hsize}{!}{\includegraphics{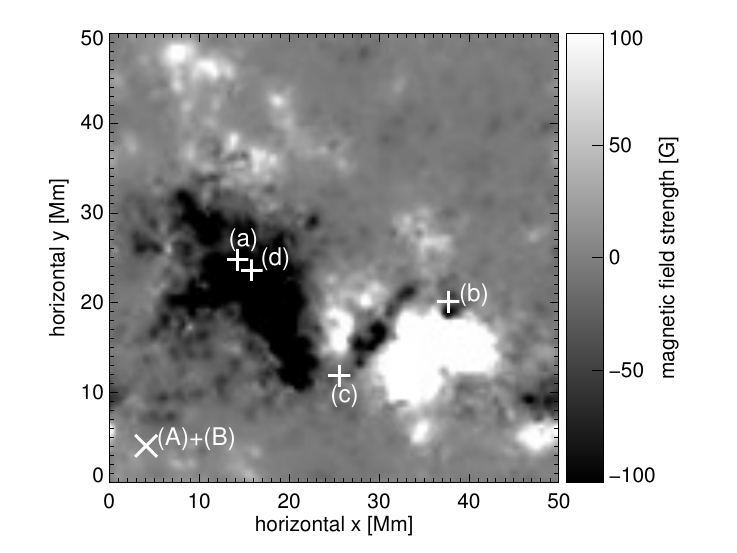}}
  \caption{Map of the vertical magnetic field component in the model photosphere
  at the lower boundary. Marked are the positions of the events
        shown in Figs.~\ref{fig:hrate_single_points} and \ref{fig:hrate_pixel_number}.}
  \label{fig:event_pos}
\end{figure}

To understand the heating of the corona, it is crucial to study its spatial
and temporal evolution. Observations only provide information about
consequences but not directly about the heating mechanism and its energy
deposition. Therefore we employ a 3D MHD model of the solar atmosphere in
order to analyze the spatial and temporal distribution of the heat
input. The data used for the present paper is based on a coronal model
introduced in \cite{2011A&A...530A.112B}. The model produces a hot corona
above a small active region with self-consistent heating. The heating in the
model is due to Ohmic dissipation of free magnetic energy created by the
braiding of magnetic field lines. This mechanism of (DC) heating has already
been proposed by \cite{1983ApJ...264..642P}, where the braiding of field
lines is a simplified picture for the energy transport due to the Poynting
flux and the subsequent dissipation process. The magnetic fields anchored in
the photosphere are shuffled around by photospheric (granular) motions
leading to a build-up of currents in the upper atmosphere. The plasma is
then heated by dissipation of these currents. Basic parameters derived from
the numerical model, such as the energy flux of roughly 100 Wm$^{-2}$ into
the corona, match observationally derived values.

The model describes the solar atmosphere above an observed active region
depicted in Fig.\ref{fig:event_pos}.  The domain spans over 50x50\,Mm$^2$ in
the horizontal direction and 30\,Mm in the vertical direction. The initial
condition for the plasma parameters are given by a hydrostatic
plane-parallel atmosphere similar to solar averages. The magnetic field is
initially computed using a potential field extrapolation.

The system is then driven by granular motions having the same statistical
properties as the photospheric velocities observed on the sun. The evolution
of the model atmosphere follows the MHD equations. The lower boundary is
prescribed by the granular motions and a fixed plasma pressure. The top
boundary is hydrostatic with no heat flux.  The magnetic field is
extrapolated by matching a potential field at the top. In the horizontal
directions the domain is fully periodic.

We ran the model for one solar hour before taking any data to make the
result independent of the initial condition. After that we took snapshots
with a cadence of 30 seconds for a bit more than a hour.

For the numerical model we employ the Pencil code
\citep{2002CoPhC.147..471B}. It solves the MHD equations for a compressible,
magnetized, viscous fluid.  These equations are the conservation of mass,
the momentum balance equation, the energy equation, and the induction
equation. The Ohmic heating term is given as $\eta \mu_0 \vec{j}^ 2$ with
$\vec{j} = \nabla\times\vec{B}/\mu_0$. The time-dependent energy balance in
the model includes anisotropic heat conduction \citep{Spitzer:1962},
optically thin radiative losses \citep{1989ApJ...338.1176C}, together with
the viscous and Ohmic heating.  No additional arbitrary heating is applied
in the model corona, which is sustained almost exclusively through the Ohmic
dissipation.

\section{Sample events: nanoflares and nanoflare storms}
\label{sec:temp_point}

\begin{figure*}
  \sidecaption
  \includegraphics[width=12cm]{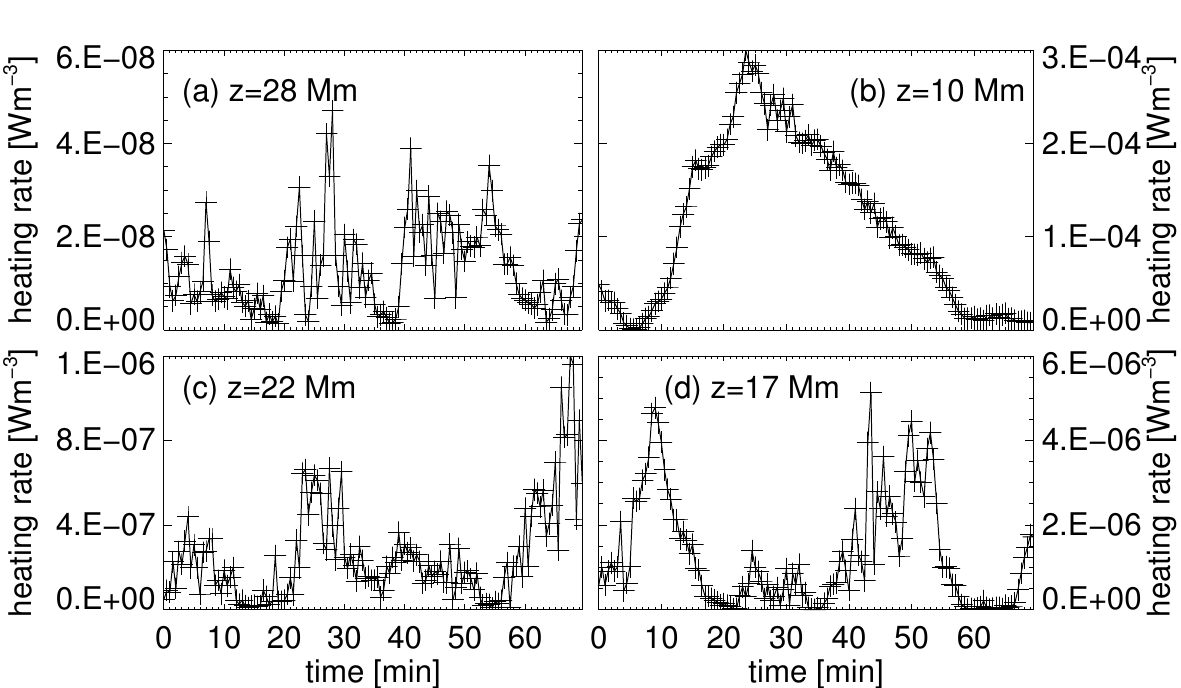}
  \caption{Temporal evolution of the Ohmic heating rate ($\sim \vec{j}^2$)
    at four different locations (single grid-points) in the model corona: (a)
    grid-point with smallest time-averaged heat input, and (b) with strongest
    time-averaged heat input in the coronal volume at $T{>}10^5$\,K; (c)
    location at the top of a bright loop seen in the synthesized coronal
    emission, and (d) at the coronal base at the footpoint of the bright
    loop.  The crosses mark the times of the snapshots.  The geometrical
    heights of these four locations are indicated in
    Fig.~\ref{fig:hrate_average}. The locations are also indicated
    in Fig.~\ref{fig:event_pos} relative to the photospheric magnetic field. } 1
  \label{fig:hrate_single_points}
\end{figure*}

The ohmic heating in the model is transient in time and space.  To
illustrate this, we show in Fig.~\ref{fig:hrate_single_points} the temporal
variation in the volumetric heating rate at four grid points of the
numerical simulation. These have been selected using the following criteria:
(a) the lowest heating rate in the box, (b) the highest heating rate in the
coronal volume, i.e., in the region where the temperature exceeds $10^6$~K,
(c) at the apex of a loop clearly visible in (synthesized) coronal emission,
and (c) at the coronal footpoint of that loop. Obviously these locations
span a wide range of (peak) heating rates, covering four orders of
magnitude.

\begin{figure}
  \resizebox{\hsize}{!}{\includegraphics{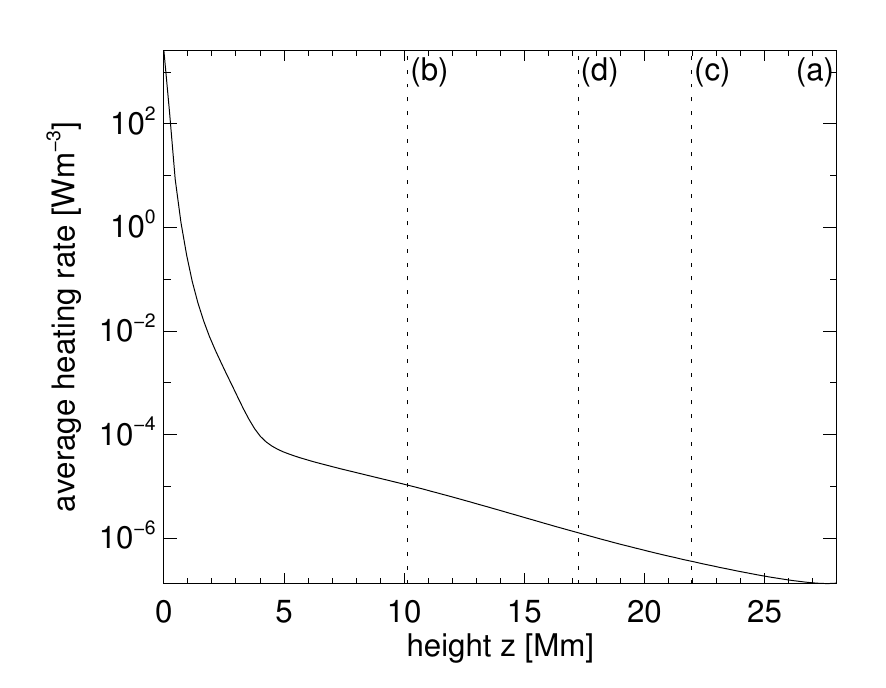}}
  \caption{Horizontally and temporally averaged Ohmic heating rate ($\sim
    \vec{j}^2$). The marked postions are at about (a) 28\,Mm, (b) 10\,Mm,
    (c) 22\,Mm, and (d) 17\,Mm according to the locations depicted in
    Fig.~\ref{fig:hrate_single_points}.}
  \label{fig:hrate_average}
\end{figure}

In part, this large difference in peak heating is due to the strong drop in
the volumetric heating rate with height. In the coronal part, the
horizontally averaged volumetric heating rate drops roughly exponentially
with a scale height of about 4 Mm, as can be seen from
Fig.~\ref{fig:hrate_average} \citep[the heating \emph{per particle} shows a
peak in the transition region; cf.][]{2011A&A...530A.112B}. For example,
this drop accounts for a difference in heating rate between locations (a)
and (b) in Fig.~\ref{fig:hrate_single_points} of about a factor of 100 (cf.\
dotted lines in Fig.~\ref{fig:hrate_average}). On top of this \emph{average}
drop with height, there is a strong spatial inhomogeneity caused by the
structure of the magnetic field and how the magnetic field lines are
braided. The magnetic field geometry above 4\,Mm \citep{2011A&A...530A.112B}
is dominated by a large loop system connecting the main polarities
(c.f. Fig.~\ref{fig:event_pos}). At heights above 4\,Mm, almost all field
lines connect to these polarities. As a result the selected positions do not
show any distinguished magnetic configuration.

Most important is that the temporal variability at each position shows a
different nature. Referring to Fig.~\ref{fig:hrate_single_points}, e.g., at
location (a) the heating is due to short pulses with a lifetime of a few
minutes, whereas at position (b) only one pulse over more than half an hour
seems to occur. At positions (c) and (d), the heating is composed of short
pulses and of pulses with lifetimes of ten or more minutes. These four
profiles represent the complete domain in the sense that the energy is
inserted in more or less randomly distributed pulses. Thus it is quite
challenging to investigate each of the million gridpoints in the numerical
domain. Thus one has to employ a statistical approach. Before we do this in
Sect.~\ref{sec:stat}, we investigate one particular location in more detail.

\begin{figure}

  \resizebox{\hsize}{!}{\includegraphics{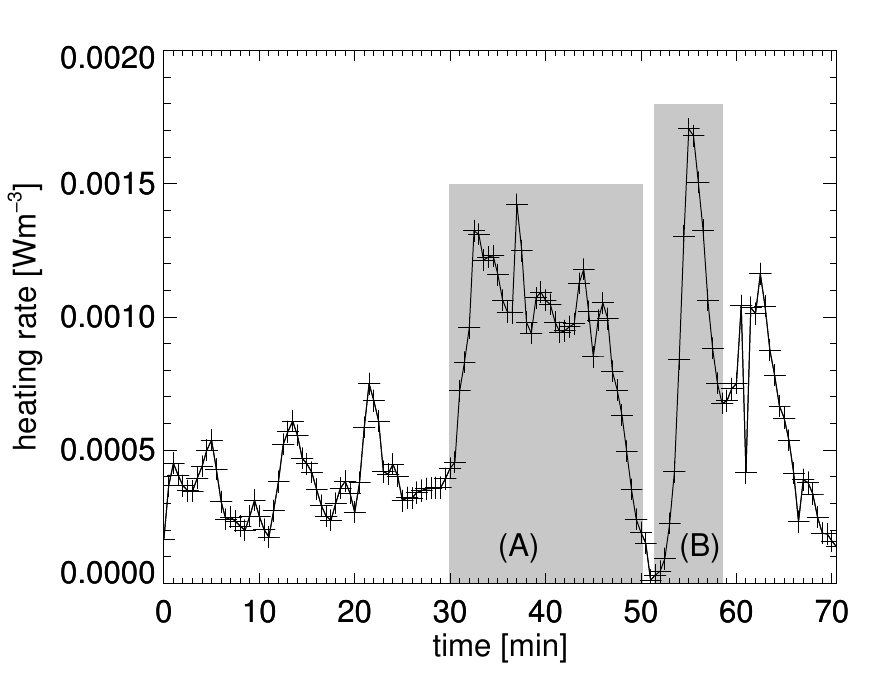}}
  \caption{Ohmic heating rate (${\sim}\vec{j}^2$) at a fixed point in space
    for more than a solar hour. The data points are marked by crosses. Here
    a gridpoint at the very base of the corona and transition region
    (${\approx}3$\,Mm height) was chosen which is representative of the
    region where the heating rate \emph{per particle} is maximal. The large
    cross in Fig.~\ref{fig:hrate_single_points} indicates the horizontal
    location. The shaded regions (A) and (B) indicate a nanoflare storm and
    a single nanoflare, see Sect.~\ref{sec:temp_point}.  }
  \label{fig:hrate_pixel_number}
%
\end{figure}

As an illustration we select one spatial location at the very base of the
corona and transition region, where the (average) heating rate \emph{per
  particle} is highest \citep[see][]{2011A&A...530A.112B}. This is shown in
Fig.~\ref{fig:hrate_pixel_number}. Looking at a spot where the heating rate
\emph{per particle} is at its peak gives some confidence that this is where
most of the action is, and thus this example can be considered a
representative one.

To estimate the energy deposited in one single burst of heating, the volume
covered by this event is needed. The current structures aligned with the
magnetic field are typically limited by the spatial resolution of the
numerical experiment. Using a high-order scheme to solve the MHD equations,
these small structures will have an extent of roughly 4x4x4 grid points
\citep[see also][their Fig.~9]{2011A&A...530A.112B}. This corresponds to a
volume of very roughly 2\,Mm$^3$ and should be considered as a lower limit.

The temporal evolution in Fig.~\ref{fig:hrate_pixel_number} shows two events
in close succession: event (A) consists of several heating pulses occurring
shortly one after another so that they merge into one single longer duration
event lasting almost 20 minutes. The second event (B) consists of a single
heating pulse with a larger amplitude lasting only a few minutes. (The same
is true for the weaker events at times 15 minutes and 22 minutes).

When integrating in time and space about 10$^{19}$\,J are deposited during
event (A) and a couple of 10$^{17}$\,J are deposited during event (B). These
energies coincide roughly with the energy of a typical nanoflare of
10$^{17}$\,J (10$^{24}$\,erg) as originally suggested from theoretical
considerations by \cite{1988ApJ...330..474P}.  Event (B) could then be
``classified'' as a regular nanoflare. In contrast, event (A) could be
related to a \emph{nanoflare storm}, a phenomenon ``with durations of
several hundred to several thousand seconds''
\citep{Viall+Klimchuk:2012}. Originally, these were introduced heuristically
to model the light curves of warm loops with the help of 1D models
\citep[e.g.][]{Warren+al:2002}, even though this name was coined only later.

Our 3D MHD model with self-consistent Ohmic heating shows that such
nanoflare storms are a natural consequence of the braiding of magnetic field
lines. Therefore our results support the ad-hoc use of such bursts for the
energy release in 1D loop models.

\section{Statistical analysis of the heating rate in time}
\label{sec:stat}

To understand the coronal heating process, it is not sufficient to study
single events alone. Instead we have to investigate statistical properties
of the heat input. For this we subdivide the computational domain into boxes
of equal size $V$, and the time span of the simulation into time intervals
of equal length $\tau$.

As one ``\emph{event}'' we define the total energy input through Ohmic
heating in one of these boxes with volume $V$ integrated over time
$\tau$. For example, if we choose $V=(5{\times}5{\times}5)$\,Mm$^3$ and
$\tau = 5$\,min, we have $10{\times}10{\times}6=600$ boxes in the
50x50x30\,Mm$^3$ domain spread over 12 bins in time for a computational time
of 60\,min. Consequently we will get $600{\times}12=7200$ individual events
for this particular combination of $V$ and $\tau$. We then investigate the
statistical properties of these events, in particular their energy
distribution. We do this for a variety of combinations of $V$ and $\tau$ to
show that our results do not depend on a particular choice of $V$ and
$\tau$. We choose the subvolumes 0.29\,Mm$^3$, 0.97\,Mm$^3$, 2.31\,Mm$^3$,
and 4.51\,Mm$^3$, which corresponds to 2$^3$, 3$^3$, 4$^3$, and 5$^3$ grid
points.  The grid spacings are d$x$=d$y$=0.39\,Mm and d$z$=0.24\,Mm.  In
this approach we use equal-sized noncubic volumes rather than searching for
the spatial extent of each individual energy deposition. In
Sect.~\ref{sec:disc_self_sim} we provide some more details on why this
approach is useful and appropriate. A decomposition in \emph{real} events is
rather challenging and perhaps even impossible. The transient heating
produces events that not only overlap in time but also overlap in
space. Additionally, the energy release shows a general decrease with
height, therefore single events will not be easily detectable, except for a
few isolated major energy releases. To reduce the investigation to those
events, which are clearly seen in the data, would reduce the number of
events dramatically, and the statistical analysis would fail.

\subsection{Energy distribution of Ohmic dissipation in the whole computational domain (from photosphere to corona)}
\label{sec:stat_self}

The procedure described above provides the energies of a large number of
events.  A common way to analyze this is to investigate the distribution of
the energies of these events.  If the frequency distribution is normalized
by the event energies $E$, we get a frequency-size distribution. In addition
we normalize the frequency-size distribution by the area and time.  Here the
area is the horizontal extent of the computational domain (50x50\,Mm$^2$),
and the time is the duration of the simulation used in this study (60\,min).

\begin{figure}
  \resizebox{\hsize}{!}{\includegraphics{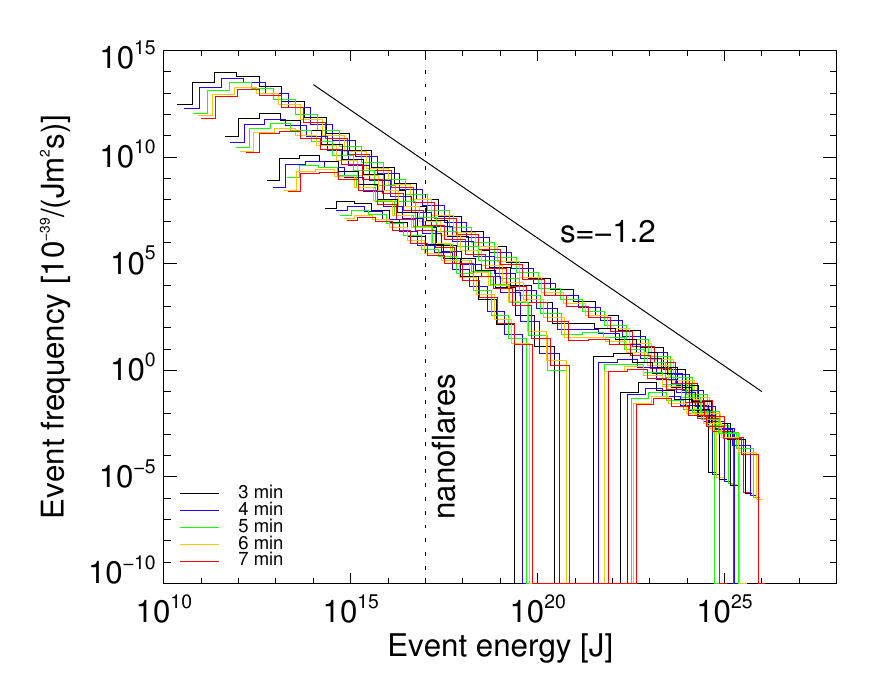}}
  \caption{Frequency distribution of Ohmic heating event energies for four
    different spatial box volumes $V$ and five different temporal bin sizes
    $\tau$. The temporal binning $\tau$ is color-coded and labeled in the
    figure. The spatial box sizes $V$ are 0.29\,Mm$^3$, 0.97\,Mm$^3$,
    2.31\,Mm$^3$, and 4.51\,Mm$^3$.  The distribution with the smallest box
    size $V$ is the group of lines starting at the smallest energies (to the
    left). For the two smallest binnings the number of events is large enough
    that the frequency size distribution is continuous. For the two
    larger binnings the distribution is interrupted owing to the low number of events.
    Overplotted is a power law with a slope of $-$1.2. The vertical dotted
    line indicates the typical nanoflare energy as predicted by
    \cite{1988ApJ...330..474P}. See Sect.~\ref{sec:stat_self}.  }
  \label{fig:pl_sl_ohmic}
\end{figure}

Figure~\ref{fig:pl_sl_ohmic} depicts the frequency-size distributions $f(E)$
of the Ohmic heating event energies for several combinations of $V$ and
$\tau$ for the \emph{whole} computational domain.  The event energies range
from 10$^{10}$\,J to 10$^{25}$\,J. These are energies several orders of
magnitude below and above the energy of a Parker nanoflare that is about
10$^{17}$\,J (10$^{24}$\,erg). The frequency distributions have a roughly
constant slope of $s\approx-1.2$ over a wide range, irrespective of the
choice of the box size $V$ or the temporal binning $\tau$.  Only for high
values of $V$ and $\tau$ are the number of events too low, and the frequency
distribution is not continuous. For instance for the largest binning in
space and time we have roughly 8000 boxes, e.g. events, in comparison to the
smallest binning with more than 10\,million boxes. If the box size
increases, small events are missing and the size of the biggest events
increases. Therefore the covered energy range of the frequency-size
distribution clearly depends on the spatial binning but depends only weakly
on the temporal binning.

As a consistency check, the frequency-size distribution can be used to
derive the total energy budget of the system. The total energy flux into the
system is given by the integral of the frequency distribution
\begin{align}\label{E:total.flux.integral}
  P  = \int_{E_1}^{E_2} f(E) ~ E ~ {\rm{d}}E \hspace{2cm} [\text{Wm}^{-2}]\,.
\end{align}
Assuming the frequency distribution to be a power law of the form $f(E) = A
E^{s}$, the energy flux in the range of energies from $E_1$ to $E_2$ is then
given by
\begin{align}\label{E:total.flux.solution}
  P  = \frac{A}{2+s}\left(E_2^{\left(2+s\right)} - E_1^{\left(2+s\right)}\right)
\end{align}
if $s \ne -2$ otherwise $P=A\ln(E_2/E_1)$. If the slope $s$ of the power law
is flatter than $-$2, then the integral is dominated by the upper energy
limit. Whereas for slopes steeper than $-$2 the result is dominated by the
lower energy limit.

The slope of the frequency distribution of Ohmic heating events is $-$1.2
and thus flatter than $-$2 (c.f., Fig.~\ref{fig:pl_sl_ohmic}).  When
considering the \emph{whole} computational domain including the photospheric
and chromospheric parts, the majority of heat is therefore deposited in
events with high energies.

The energy flux into our modeled box derived from the frequency distribution
in Fig.~\ref{fig:pl_sl_ohmic} based on Eq.~(\ref{E:total.flux.integral}) or
(\ref{E:total.flux.solution}) is on the order of $P\approx10^6$\,W\,m$^{-2}$
(${=}10^9$\,erg\,s$^{-1}$\,cm$^{-2}$). This is the flux at the photospheric
level, which is the lower boundary of our model atmosphere. This represents
the average flux over roughly one hour solar time. It matches nicely with
\cite{2011A&A...530A.112B}, where the energy flux input into the
computational domain at a fixed time was found to be in the range between
10$^6$\,W\,m$^{-2}$ and 10$^7$\,W\,m$^{-2}$.

It should be emphasized that these results apply to the \emph{whole}
computational domain, including the photosphere and chromosphere. In
Sect.~\ref{sec:stat_cor} we discuss the energy distribution and budget for
the coronal part of the domain alone.

\subsection{Power-law test for energy distribution}
\label{sec:power.test}

Even if the slope of the distribution functions of the Ohmic heating events
looks fairly constant over several orders of magnitude, we briefly discuss
how to test the validity of the assumption that this is indeed a power law.
It can be problematic to use linear regression in a log-log plot to retrieve
the slope of a power law. Therefore we follow \cite{Virkar+Clauset:2012} and
use a maximum-likelihood estimation (MLE) to obtain the slope of a possible
power law. As discussed in \cite{Virkar+Clauset:2012}, the result is
sensitive to the selection of the lower energy cutoff.  In
Table~\ref{ta:MLE_slopes} we list the estimated slopes following from the
MLE technique for various combinations of the spatial box sizes $V$ and time
bins $\tau$ for the events. We neglect 5\% of the lowest energies from the
analysis.

Having derived a slope is not yet a proof for a power-law. Again we follow
\cite{Virkar+Clauset:2012} and use the Kolmogorov-Smirnov test
\citep{Press+al:1992} to check whether the distribution functions match the
power laws. The slopes retrieved by the MLE give a confidence level above
0.9 for the two distribution functions with smallest spatial binnings (c.f.,
Fig.~\ref{fig:pl_sl_ohmic}). The highest confidence levels we find are for a
slope of 1.20. All slopes discussed here are around 1.2, and the confidence
levels are high enough to argue that the distributions of Ohmic heating
events are consistent with power laws.

\begin{table}
  \caption{Slopes of the distributions shown in Fig.~\ref{fig:pl_sl_ohmic} using maximum likelihood estimations. See Sect.~\ref{sec:power.test}.}
\label{ta:MLE_slopes}
\centering
\begin{tabular}{l r | l l l l l}
\hline
\hline
& $\tau$ [min] & 3 & 4 & 5 & 6 & 7 \\
 $V$ [Mm$^3$] & & & & & \\
\hline
0.29 && 1.19 & 1.19 & 1.19 & 1.20 & 1.20 \\
0.97 && 1.19 & 1.19 & 1.19 & 1.19 & 1.19 \\
2.31 && 1.16 & 1.16 & 1.16 & 1.16 & 1.17 \\
4.51 && 1.12 & 1.12 & 1.12 & 1.12 & 1.12 \\
\hline
\end{tabular}
\end{table}

\subsection{Heating statistics for the coronal part alone}
\label{sec:stat_cor}

We now focus on the coronal part. We define the coronal part either through
a threshold in temperature or as the volume above a certain height. Both
definitions (when properly chosen) give similar results. This will give us
information about the distribution of the heat input into the model
corona. Therefore, we repeat the analysis from Sect.~\ref{sec:stat_self} in
the subvolume of the computational domain representing the corona and the
transition region.

To define the coronal volume, in the first case we choose a temperature
threshold of log\,$T$/[K]=3.9 to include everything above the
chromosphere. In the second case we define the coronal volume as being at
heights above 4\,Mm. This corresponds to the height where the horizontally
averaged temperature exceeds $\log T/[K]=3.9$ in our model. It is also the
height where the scale height of the horizontally averaged volumetric
heating rate increases and becomes constant for the coronal part \citep[see
Fig.~\ref{fig:hrate_average} and][]{2011A&A...530A.112B}.  The model
atmosphere is very dynamic so that the isosurface for a constant temperature
is highly corrugated. The height variation of the isothermal surface of
$\log T/[K]=3.9$ is several Mm. This is larger than to the
\emph{chromospheric} scale height, but much smaller than the \emph{coronal}
scale height. In particular, this is much smaller than the vertical extent
of the coronal part of our computational domain. This is the basic reason,
why the results for these two methods will be essentially the same.

\begin{figure}
   \resizebox{\hsize}{!}{\includegraphics{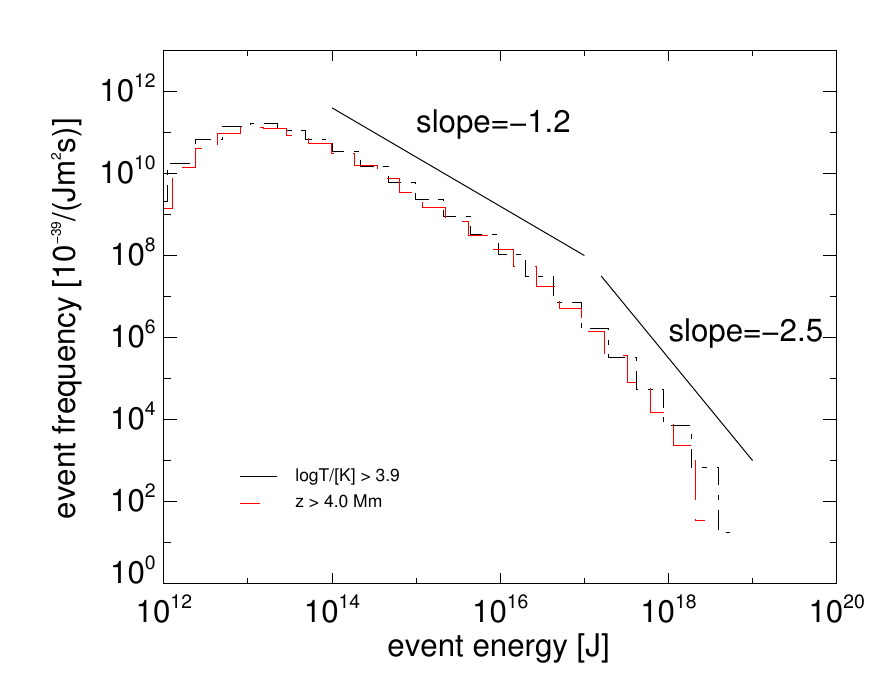}}
   \caption{Distribution of Ohmic heating event energies in the coronal
     volume alone. The results are shown for two definitions of the coronal
     volume: (a) log$T$[K]$>$3.9 and (b) heights $z{>}4$\,[Mm]. Results are
     shown only for $V{=}0.29$\,Mm$^3$, $\tau{=}5$\,min). Overplotted are
     lines indicating power laws with slopes of $-$1.2 and $-$2.5,
     respectively. See Sect.~\ref{sec:stat_cor}.  }
  \label{fig:pl_sl_ohmic_T_z}
\end{figure}

When now considering the coronal volume alone (using the above two
definitions), the event energies show a distribution that is composed of two
power laws, with a knee at about $10^{17}$\,J, and a slope of $-1.2$ below
and $-2.5$ above the knee (see Fig.~\ref{fig:pl_sl_ohmic_T_z}).
The slope at small-event energies is comparable to
Fig.~\ref{fig:pl_sl_ohmic} as found for the \emph{whole} computational
domain. For energies below 10$^{14}$\,J the distribution shows the typical
cut off due to limited resolution and statistics.
Here and in the following we only show the results for one particular choice
of the spatial box size $V$ and temporal binning $\tau$ to determine the
distribution of event energies ($V{=}0.29$\,Mm$^3$, $\tau{=}5$\,min). This
is only to not clutter the plots, the results are independent of the
particular choice of $V$ and $\tau$ (c.f. Sec.~\ref{sec:disc_self_sim}). The
distribution in Fig.~\ref{fig:pl_sl_ohmic_T_z} represents a subsample of the
complete computational domain so that the range of the covered event
energies is smaller than for the analysis of the entire domain shown in
Fig.~\ref{fig:pl_sl_ohmic}.

The result that the slope of the power law is flatter than $-2$ below
$10^{17}$\,J and steeper than $-2$ above has a very pointed
consequence. Below the knee most of the energy is deposited at higher
energies, i.e., close to the knee. Above the knee most of the deposition is
at low energies, i.e., again close to the knee. Therefore most of the energy
will be deposited close to the knee at $10^{17}$\,J, which coincides with
the originally proposed energy for nanoflares.

\begin{figure}
\resizebox{\hsize}{!}{\includegraphics{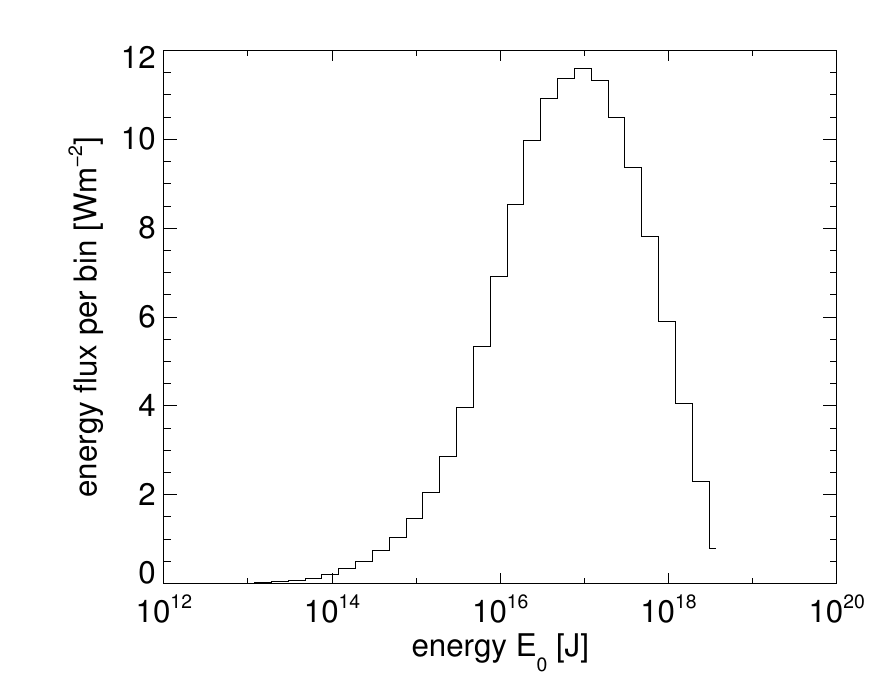}}
\caption{Energy fluxes into the corona deposited at different energies $E_0$
  of the heating events. Derived from the distribution in
  Fig.~\ref{fig:pl_sl_ohmic_T_z}. The distribution is first interpolated to
  allow for an integration interval of 0.2\,dex in logarithmic values.  See
  Sect.~\ref{sec:stat_cor}. }
  \label{fig:pl_T_int_b}
\end{figure}

A more \emph{quantitive} way to locate the main contribution to the heat
input in terms of event energies is to evaluate the energy deposited in
different energy ranges. For this, we integrate the frequency distribution
in Fig.~\ref{fig:pl_sl_ohmic_T_z} piecewise with a fixed interval in
$\Delta\log E\,[J]= 0.2$ around the energy $E_0$
\begin{align}
  P(E_0) = \int_{E_0-\Delta E}^{E_0+\Delta E} f({E})~{E}~\text{d} {E} ~.
\end{align}
This gives the total energy flux into the corona that is deposited in a
range of event energies centered on $E_0$.  Figure~\ref{fig:pl_T_int_b}
shows the distribution of there energy fluxes $P(E_0)$ as a function of the
event energy $E_0$. The maximum flux is found at an energy of roughly
10$^{17}$\,J (10$^{24}$\,erg) just as suspected above.  The integration is
performed using an interpolation of initial distribution. But using the
original resolution in energy does not alter the results, and only the width
of the bins would be broader. The error for the peak flux is estimated by
the half bin width of initial data, e.g, 0.16 or a factor of roughly 1.4

%

As a sanity check, we can use the result shown in Fig.~\ref{fig:pl_T_int_b}
to compute the total energy input to the model corona by integrating over
the whole range of energies. We find this to be 100\,Wm$^{-2}$, which is
consistent with the results given by \cite{2011A&A...530A.112B} and
consistent with estimations of the coronal heating
requirements based on observations \citep{1977ARA&A..15..363W}.

This result of the distribution of energy fluxes in
Fig.~\ref{fig:pl_T_int_b} shows that the upper atmosphere in our model is
mainly heated by events with an energy of a few times 10$^{17}$\,J, which
roughly matches the typical nanoflare energy. In addition to the
\emph{qualitative} discussion of the position of the knee in
Fig.~\ref{fig:pl_sl_ohmic_T_z}, this \emph{quantitative} result also
provides information on the \emph{range} of energies responsible for the
energy input. From Fig.~\ref{fig:pl_T_int_b} it is clear that energies from
10$^{16}$ to 10$^{18}$\,J (10$^{23}$ to 10$^{25}$\,erg) contribute
significantly.

From this we can conclude that in our 3D MHD model there is a preferred
energy for the deposition through Ohmic heating. This value is the same
order of magnitude as the nanoflare energies derived by
\cite{1988ApJ...330..474P} on theoretical grounds, despite all the
limitations in terms of spatial resolution and magnetic Reynolds number in
our numerical experiments. This is reassuring, because our 3D MHD model
basically describes the flux-braiding process as proposed by
\cite{1988ApJ...330..474P}.

\section{Consequences for observable quantities} \label{sec:obs}

The frequency-size distributions derived in the previous section are based
on the energy deposition due to Ohmic dissipation. However, these events
with energies close to the values of nanoflare energies are not directly
observable on the Sun. For an observational study of the region where the
plasma is heated, the only source of information we have is the
\emph{emission} in exteme UV and X-rays. And this provides only partial
information on the energy budget, because heat conduction, particle
acceleration, and maybe other processes will also redistribute the energy,
which will then be, e.g., radiated in other places.

To study how the distribution of energies changes if one investigates the
brightenings related to the heating events, we synthesize the expected
optically thin coronal and transition region emission from the MHD
model. Here we follow the procedure outlined by
\cite{Peter+al:2004,Peter+al:2006}. To derive reliable estimates of the
coronal emission, the MHD model has to treat the energy equation
appropriately; in particular it is essential to include the heat conduction
to self-consistently set the proper coronal pressure (and thus
density). Using the CHIANTI atomic data base
\citep{Dere+al:1997,Landi+al:2006-chianti}, we calculate the emission at each grid
point for a number of emission lines based on the temperature and density at
the respective gridpoint in the model. The result is the volumetric energy
\emph{loss} per second through optically thin radiation in a given spectral
line.

\begin{figure}
  \resizebox{\hsize}{!}{\includegraphics{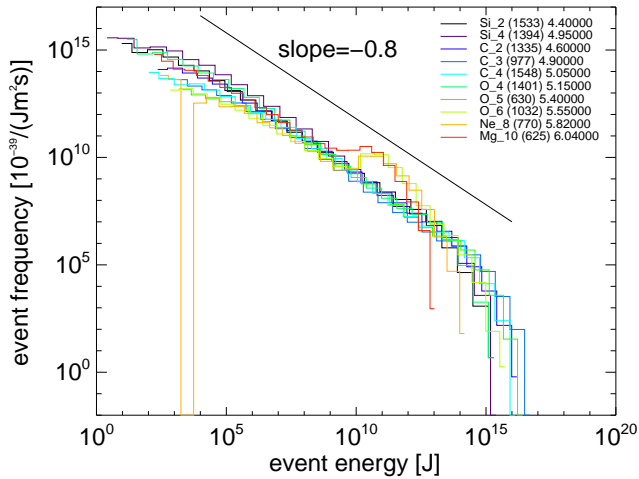}}
  \caption{Event frequency distributions of synthesized emission lines. The
    name of the lines are given in the legend along with the wavelength
    [\AA] and the line formation temperature in $\log T/[\text{K}]$.}
  \label{fig:pl_sl_em2}
\end{figure}

We repeat the analysis of the preceding section and treat the emission in
each of the lines in the same way as the energy \emph{input} in
Sects.~\ref{sec:stat_self} and
\ref{sec:stat_cor}. Figure~\ref{fig:pl_sl_em2} shows the resulting
distribution functions for the radiative energy loss events. The energy range is
shifted to much lower energies than in Figs.~\ref{fig:pl_sl_ohmic}
and \ref{fig:pl_sl_ohmic_T_z}. This is mainly because each line only
represents one single emission line, and each emission line contributes
only a small fraction of the total energy loss.  Nonetheless, the
distribution functions show similar power laws for all emission lines with
line formation temperatures ranging from 30\,000\,K up to 10$^6$\,K. Owing to
the different formation temperatures, the lines originate from different
densities, so that the frequency distributions have different cut-offs at the
high-energy end.

Most importantly, the power law shows a slope of about $-0.8$. This slope is
slightly flatter at low energies and a bit steeper for high energies, but is
flatter than the slope of $-1.2$ found for the energy input.  The reduction
of the slope reflects the nonlinear connection from heat input to the final
(optically thin) radiative output. There are basically two ways of looking at
this phenomenon: (1) a heuristic explanation based on the energy
redistribution and (2) a more formal argument using scaling laws.

At coronal temperatures heat conduction dominates radiative losses
\citep[e.g.][]{Priest:1982}; the higher the temperature, the more heat
conduction dominates (below $10^7$\,K). \emph{On average}, the heat input
(per volume and time) is smaller at higher temperatures at larger heights
(Fig.~\ref{fig:hrate_average}), where the energy input is lower. Thus heat
conduction is the major loss process at the low-energy end of the
distribution of event energies. Consequently, there is less radiation at the
low-energy end, and the distribution of the radiative losses will be
flatter, in our case the slope is only $-0.8$ instead of $-1.2$.

The same mechanism also explains why the knee in the energy input
(Fig.~\ref{fig:pl_sl_ohmic_T_z}; Sect.~\ref{sec:stat_cor}) is not clearly
visible in the distribution of the radiative losses
(Fig.~\ref{fig:pl_sl_em2}). The energy deposited around the nanoflare
energies coinciding with the location of the knee is mainly transported down
towards the chromosphere, i.e.\ towards the high-energy end of the
distribution, and is then radiated at lower temperatures in the transition
region. In turn, this enhances the distribution at the high-energy end of
the distribution of the radiative output.

To get a quantitative estimate of the flattening of the power law slope from
the energy input ($-1.2$) to the radiative output ($-0.8$), we investigate
the scaling laws as already described by \cite{Rosner+al:1978}. The scaling
laws relate the volumetric heat input $E_H$ (assumed to be constant) and the
loop length $L$ to the peak temperature $T$ and pressure $p$ of a
hydrostatic loop \citep[Eq. 4.3 and 4.4 of][]{Rosner+al:1978}. One can
easily combine these to relate the density $n$ at the apex to the heat input
and loop length, $n \propto E_H^{4/7} L^{1/7}$. Because the dependence on
$L$ is weak, we discard this in the following. The energy loss $\varepsilon$
due to optically thin radiation (per volume and time) is proportional to the
density squared, $\varepsilon \propto n^2$, and with the above scaling law
we find $\varepsilon \propto (E_H)^{8/7}$. Therefore the slope of the power
law should change from $-1.2$ to $-1.2{\times}7/8=-1.05$. This change does
go in the right direction, but does not quite explain the result of $-0.8$
we find in Fig.\ref{fig:pl_sl_em2}. This is not very surprising, because the
\cite{Rosner+al:1978} scaling laws apply only to hydrostatic loops with a
constant heating rate, while our 3D MHD model is time-dependent with a
nonconstant heating rate.

In conclusion, the energy events for the radiative \emph{losses} are still
distributed as a power law, but the slope is flatter than the power law for
the actual energy \emph{input}. This is important for observational studies
investigating the energy input by analyzing the energy loss through
brightenings seen in EUV and X-rays.

\section{Self-organized criticality and fractal dimension}
\label{sec:disc_self_sim}
In Sect.~\ref{sec:stat} we derived the frequency-size distributions for Ohmic
heating events. These distributions follow a power law over several orders
of magnitude (c.f. Fig.~\ref{fig:pl_sl_ohmic}). This power law does not
change if the binning of the data cube is changed, either spatially or
temporally.  This behavior is typical of a power-law frequency-size
distribution, so it indicates that our modeled energy input also follows a power law.
Systems that are characterized by a power law are called
self-organized critical systems \citep{Bak+al:1987}.  Self-organized
criticality can also be applied to other systems such as $1/f$ noise, and it
shows the scale invariance of the system on a macroscopic level. This scale
invariance manifests itself by the fact that the slopes of the power laws
are independent of the spatial scale range under consideration, i.e. of the
spatial binning.  Therefore the approach of using a constant binning for the
analysis in Sects.~\ref{sec:stat_self} and \ref{sec:stat_cor} does not
affect the results.
The constant  binning refers to equally sized volumes. Likewise the noncubic
boxes do not alter the result since we can use arbitrary volumes in
a scale-independent system.

We find in our analysis that the frequency size distribution of the event
energies can be represented by a single slope alone.  This is consistent
with other model approaches, e.g, \cite{Dimitropoulou+al:2011} and
\cite{Gerogoulis+al:2001}, and with observations \citep{Aschwanden+al:2012}.
In earlier studies \cite{Georgoulis&Vlahos:1996} and
\cite{Georgoulis&Vlahos:1998} used cellular automata models and find a
double power law behavior for the event size.  That our
distribution functions follow a single power law can be accounted for by the
short time period of one hour and by the small field of few of our model active
region.

Following \cite{2010arXiv1008.0873A} the slope of the power law of the
distribution functions point to the propagation properties of the
instability leading to the dissipative events. The fractal dimension $D$ is
thereby correlated to the slope $s$ by $D=3/s$
\citep{2010arXiv1008.0873A,Aschwanden+al:2012}. Since we measure a slope of
$1.2$, the fractal dimension of the system would be 2.5 for the Ohmic
dissipation events. This means that the fractal dimension is not bound to
the 2D surface of separatrix layers, and that the propagation
of the instability cannot spread freely in all three dimensions.

Formally, using the slope of $-1$ of the power law for the radiative losses
(c.f. Fig.~\ref{fig:pl_sl_em2}), we find a fractal dimension of 3. However,
because the emission is just a secondary quantity following from the heat
input and strongly affected by the heat conduction, the diagnostic value of
this result is limited, though.

\section{Conclusions}

In Sec.~\ref{sec:temp_point} we discussed the transient nature of the
heating based on time series of the Ohmic heating at specific locations.
There the temporal evolution of the heating rate shows a stochastic nature,
as well as a qualitative difference at different places.  We saw long bursts
(10 min and longer), which may consist of several short pulses. At other
locations we saw only short pulses (1 min and less).  These events could
correspond to the nanoflare storms and nanoflares often assumed in 1D loop
models \citep{Viall+Klimchuk:2012}.  We used these time series only for a
qualitative description and for a rough estimate of the magnitude of heating
events. We saw that these events deposit energies on the the order of
10$^{17}$\,J, which agrees with the nanoflare energy estimated by
\cite{1988ApJ...330..474P}.

We also presented a more sophisticated approach using statistical methods to
analyze these events in the complete domain.  The statistical analysis of
the coronal heating reveals a power law for the heating event energies. Even
though the driver, i.e. the granular motion, does not work on all scales,
the slope of the power law is constant over several orders of magnitude and
flatter than $-$2. We find that in the transition region and corona the
dominant energy for the heating events is the nanoflare energy,
i.e. 10$^{17}$\,J.  Since the higher energy events occur on average at lower
heights, the heating is footpoint-dominated. The event energies we find in
our model are below the detection limit of current instrumentation and
support the scenario of nanoflare heating as proposed by
\cite{1988ApJ...330..474P}.

In our model the energy provided by the field-line braiding is converted
into heating through Ohmic dissipation and not into particle
acceleration. This implies that non local effects are not incorporated. This
is justified by the electron mean-free-path being smaller than the grid
spacing of the model. The energy is redistributed by advection and
conduction or radiated by optically thin emission. From our model atmosphere
we synthesized observable extreme UV emission lines. The statistical
analysis of these lines shows again a power-law behavior. The slope of this
power law is around 1 and smaller than the slope of the heating
process. Even though the power law is maintained over several orders of
magnitude, in contrast to the heating process, the radiative losses are not
in a self-organized critical state. This is because the radiative losses are
coupled to the heating in a nonlinearly.

In this study we showed that a 3D MHD model of the active region solar
corona has a preferred energy for depositing of heat through
Ohmic dissipation following field-line braiding. The preferred energy
of about 10$^{17}$\,J (10$^{24}$\,erg) corresponds well to the value
derived by
\cite{1988ApJ...330..474P} and thus is yet another piece of evidence
that, despite all their shortcomings, the 3D MHD models for the flux-braiding
mechanism provide a good description of the heating of the solar corona
in active regions.

\bibliographystyle{aa} \bibliography{literature,lit_new}

\end{document}